# Learning through experimenting: an original way of teaching geometrical optics

C. Even[1], C. Balland[2] and V. Guillet[3]

[1] Laboratoire de Physique des Solides, CNRS, Université Paris-Sud, Université Paris-Saclay, F-91405 Orsay, France

[2] Sorbonne Universités, UPMC Université Paris 6, Université Paris Diderot, Sorbonne Paris Cite, CNRS-IN2P3, Laboratoire de Physique Nucléaire et des Hautes Energies (LPNHE), 4 place Jussieu, F-75252, Paris Cedex 5, France

[3] Institut d'Astrophysique Spatiale, CNRS, Université Paris-Sud, Université Paris-Saclay, F-91405 Orsay, France

E-mail : catherine.even@u-psud.fr

## Abstract

Over the past 10 years, we have developed at University Paris-Sud a first year course on geometrical optics centered on experimentation. In contrast with the traditional top-down learning structure usually applied at university, in which practical sessions are often a mere verification of the laws taught during preceding lectures, this course promotes "active learning" and focuses on experiments made by the students. Interaction among students and self questioning is strongly encouraged and practicing comes first, before any theoretical knowledge. Through a series of concrete examples, the present paper describes the philosophy underlying the teaching in this course. We demonstrate that not only geometrical optics can be taught through experiments, but also that it can serve as a useful introduction to experimental physics. Feedback over the last ten years shows that our approach succeeds in helping students to learn better and acquire motivation and autonomy. This approach can easily be applied to other fields of physics.

## 1. Introduction

Teaching geometrical optics to first year university students is somewhat challenging, as this field of physics often appears as one of the less appealing to students. Indeed, geometrical optics is often considered as an old-fashioned and dusty science. This is in part due to the fact that the main laws governing this discipline were discovered centuries ago (e.g., Snell's laws) and are usually taught in a rather formal way. This certainly also pertains to the fact that technological advances in this field and their impact on today's modern societies are less obvious than in other fields of Physics, e.g. electronics or opto-electronics. Moreover, it is common that the teaching of geometrical optics is not accompanied with practical sessions.

Even when those do exist, practical works in geometrical optics often appear less sophisticated than practical works in other fields, as the main detector is the human eye and only rather 'simple' elements are involved.

In France, this feeling with respect to geometrical optics might be exacerbated by the rather formal way physics is traditionally taught. Teaching at university is indeed structured as a top-down process: the knowledge is delivered by the professors during lectures in a somewhat monolithic way, tutorials are formatted for the student to assimilate the concepts of the course and practical sessions, when they exist, are often a mere verification of the physical laws stated during the lectures. One major drawback of teaching first year students this way is that this structure does not adapt to the vastly heterogeneous levels of students arriving at university. Moreover, the multiplication of courses, tutorials and practical sessions yields a very dense schedule: little time is left for the student to integrate the new concepts and acquire autonomy. Finally, due to the competition with the French 'Grandes Ecoles' (selective and prestigious engineering schools whose curriculum is generally considered of a higher academic level) students arriving at university often feel they are 'second-class' students and strongly lack motivation for study.

As a matter of fact, geometrical optics is paradoxically a science with strong assets for introducing experimental physics at University. Experimental protocols are simple to implement (at least in the first years of University), measurements are quite easy to make, results are visual and often spectacular, uncertainties are non negligible and rather easy to grasp. Hence, geometrical optics turns out to be particularly suitable to be taught through experiments. Moreover, for all the reasons outlined above, it makes sense to introduce experimental physics through geometrical optics.

In this spirit, our team at University Paris Sud Orsay decided a decade ago to modernize the ways of teaching geometrical optics to first-year students by breaking the traditional lectures-tutorials-practicals trio by putting practical works at the heart of the learning process. The course's pedagogical goal then is not so much to provide concepts but rather to teach experimental skills that will be useful in other courses of experimental physics such as mechanics or wave physics. In this spirit, the course focuses on measurements rather than on a comprehensive description of optical devices. For example, while several optical instruments are studied during tutorial sessions, a whole practical session is devoted to a single instrument: the telephoto lens. The purpose we follow in this course is three-fold: motivate the students by making them more active, smooth the transition from high school to university and open the teaching towards modernity through everyday life and applications. Some aspects of our approach are borrowed from the 'flipped classroom/active' pedagogy that was developed during the nineties in the US and Canada (e.g., [1]), and more generally from the philosophy of active learning (see, e.g., [2,3] for a general approach to active learning or, e.g., [4-6] for active learning applied specifically to optics), in particular learning through discovery and developing the students autonomy by encouraging them to solve the problems they face by themselves (see, e.g., [7] for an introduction of active learning into French university physics classrooms).

In this paper, we first present the main principles of our new approach of teaching geometrical optics through experiment (section 2). We focus on four main notions of the course to illustrate the running of a typical session. We discuss in section 3 the advantages and drawbacks of this approach. Section 4 presents the accompanying materials that have been developed for the course. The organization of the course and the evaluation of students are presented in sections 5 and 6, while feedback on our work and possible improvements are discussed in sections 7 and 8. The curriculum of the course is presented in an appendix.

## 2. Experiment at the heart of teaching

In an ordinary course, the practical part is usually considered (by the students but sometimes even by the professors) as an application, a check of the theory exposed by the professor during his/her lecture. Therefore, the students sometimes don't appreciate or don't see the interest of experimenting. But one should not forget that physics is an experimental science and that many laws were first discovered experimentally. This is the case, for instance, of the so-called "Snell's laws" (which are named "Descartes's laws" in France…) which were first discovered by Ptolemy in the Antiquity using a disk immersed in water.

In order to follow this historical approach, but not only, the experiment is done first in the teaching course presented here. More precisely, the presentation of a phenomenon or concept requires three steps:

1. The student first discovers the phenomenon through qualitative experiments that are mainly made by the students, and in some cases by the professor
2. The underlying physical concepts are then exposed by the professor
3. The student comes back to the experiment, checks the adequacy of his/her first conclusions to the course results, then makes quantitative measurements (with uncertainty estimates) to strengthen his/her knowledge

This process is repeated for each new concept under investigation during a learning session. During the experiments (steps 1 and 3), the students are encouraged to interact with others to share questions and findings, thus following the idea of "peer instruction" [8].

In this structure, each physical concept under investigation is thus associated with a set of qualitative and quantitative experiments. In order to illustrate the above process, we present four fundamental concepts that belong to the set of competences the student has to acquire in this course: Snell's laws, total internal reflection, virtual image formation, focal distance of a thin lens. For each of the above, the following table summarizes the corresponding set of experiments.

| Concept | Qualitative experiments | Quantitative experiments |
|---|---|---|
| Snell's laws | Reflection and refraction of a laser beam at the surface of water | Measurement of incident ($i$), reflected and refracted ($r$) angles |

| | Reflection and refraction of a laser beam on a Plexiglas hemi-cylinder | Graph of *sin i* as a function of *sin r* to deduce Snell's law, from which the refraction index of Plexiglas is deduced |
|---|---|---|
| Total internal reflection | Observation of successive internal reflection within a falling water stream, a Plexiglas ruler, and an optical fiber | Measurement of the critical angle for total internal reflection, from which the refraction index of Plexiglas is deduced |
| Virtual image | Observation and characterization of the image formed by a plane mirror. | Quantitative determination of the position of the images. Graphical representation of the formation of the images |
| Focal distance of a thin lens | Observation of the image of a vertical slit formed by a thin convergent lens as a function of the lens-object distance. | Measurement of focal distances with three different methods:<br>- materialization as the convergence point of an incident parallel beam;<br>- auto-collimation method;<br>- graph of the inverse of the image position as a function of the inverse of the object position |

*a) Introducing Snell's laws*

<u>*Step 1:*</u> To introduce Snell's laws, a first qualitative experiment is conducted by the teacher using a recipient filled with water, which shows both reflection and refraction of a laser beam at the air-water interface. Then, the students observe the phenomenon by themselves using a purpose-built setup: the classical Plexiglas half-cylinder attached to a graduated disk and a laser beam.

<u>*Step 2:*</u> After these simple observations, Snell's laws are presented by the teacher in a formal way and put into historical perspective.

<u>*Step 3:*</u> Students then make the measurements of both the incident and reflected/ refracted angles. They observe that the reflected angle is equal to the incident one *i*, and they trace *sini* as a function of *sinr* (*r* is the refracted angle) in order to deduce Snell's law. Using this curve, they also deduce the refraction index of Plexiglas.

*b) Total internal reflection*

<u>*Step 1:*</u> A second example is the study of total internal reflection. A thin He-Ne laser beam is sent by the teacher in the direction of an opening at the bottom of a recipient filled with water. The water pours out of the recipient and forms a falling water stream in which the light undergoes several successive internal reflections (see Fig 1a). Then, the students observe by themselves the successive reflections inside a Plexiglas ruler (Fig 1b) and in an optical fiber (Fig 1c). During these experiments, they are confronted with several key phenomena to understand the guided propagation of light: coupling issues, losses, attenuation.

_Step 2:_ The professor then presents the formal application of the law of refraction to total internal reflection. The existence and the expression of the critical angle for total internal reflection is derived.

_Step 3:_ The students finally measure the critical angle of total reflection using the Plexiglas hemi-cylinder they used previously for determining Snell's law. From this measurement, they deduce the index of the Plexiglas.

At the end of the study of total internal reflection, the students are able to compare the two measurements of the refraction index of Plexiglas with their uncertainties (see section 4b).

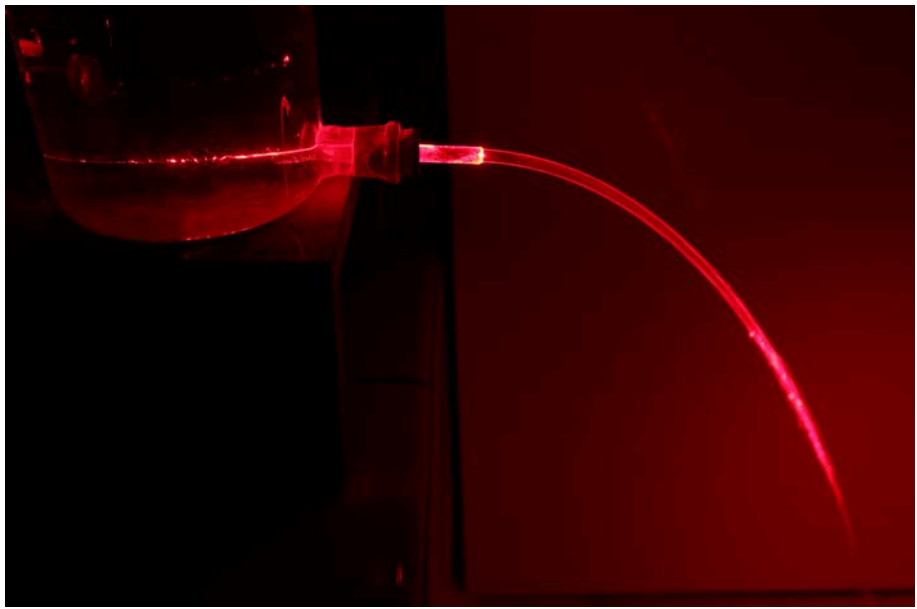

**a)**

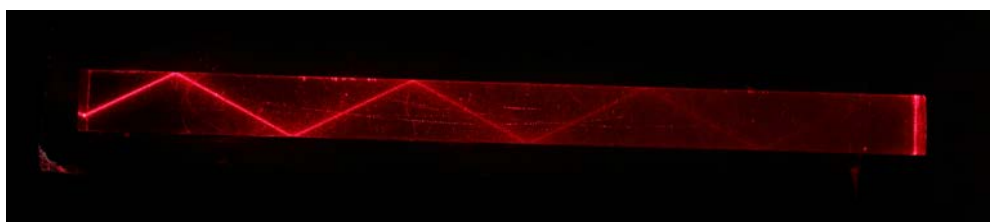

**b)**

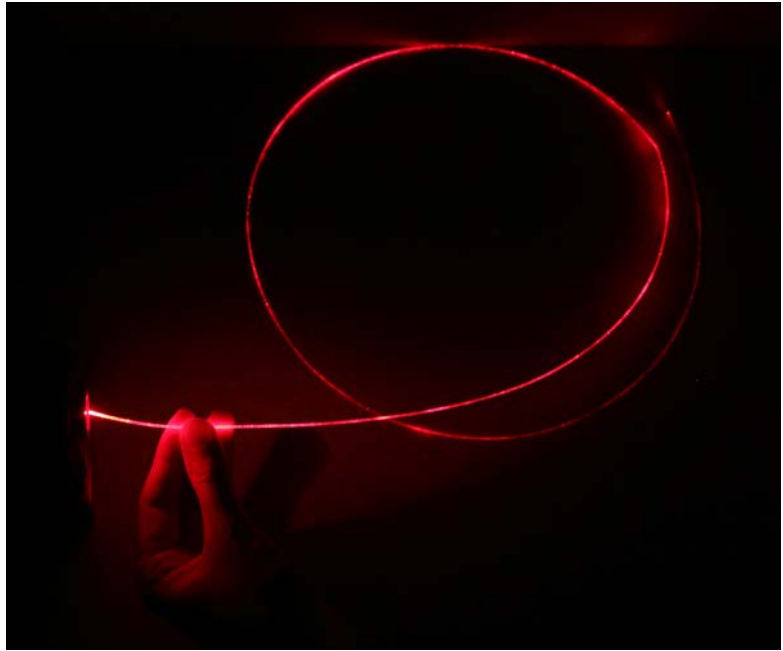

**c)**

**Fig 1: guiding light with total internal reflection : a) falling water stream ; b) Plexiglas ruler ; c) optical fiber**

*c) Virtual image obtained with a plane mirror*

The notion of virtual image produced by an optical system is often presented in a rather academic and abstract way. The optical system divides space into two parts: the object space and the image space. An image forming in the image space is said to be real, whereas an image forming in the object space is said to be virtual. From this definition, students have difficulty in understanding the notion of virtual image: for them, its virtual nature implies that such an image cannot be observed. They do not associate their every day experience of looking at themselves through a mirror with seeing a virtual image.

*Step1*: In the present course, the notion of virtual image is introduced experimentally. A small light bulb (light on) is placed in front of a Plexiglas plate and both the bulb and its reflection on the plate are visible. Students seek the apparent position of the reflected bulb by placing an identical bulb (off) behind the plate and by trying to match its position with the reflection of the on light bulb (see Fig 2a). They find that the reflected light seems to originate from a position behind the plate, symmetric to the real bulb. The Plexiglas plate is then replaced by a plane mirror: it becomes obvious that the reflected image is located behind the mirror, is symmetric from the object with respect to the mirror (Fig 2b) and, although it is perceivable by eye, no light rays come from behind the mirror plane.

*Step 2:* After this series of experiments, the professor defines the notions of virtual or real images from an observational point of view: a real image can be observed on a screen where light rays converge. A virtual image cannot be projected on a screen but it can be seen with the eye, the light rays do not actually meet, only their backward extrapolations through the optical system intersect.

*Step 3:* On a graph, an object and a plane mirror are sketched. The students trace three light rays originating from the top of the object with different directions and apply Snell's second law on the mirror to follow their trajectories after reflection. They do the same with three light rays emanating from the bottom of the object. They then seek the locus of intersection of each set of rays and find that the backward extensions of the rays meet beyond the plane of the mirror. They construct the image and find that it is behind the mirror symmetrically to the object, straight and of the same size of the object. They check the consistency of these findings with their previous observations.

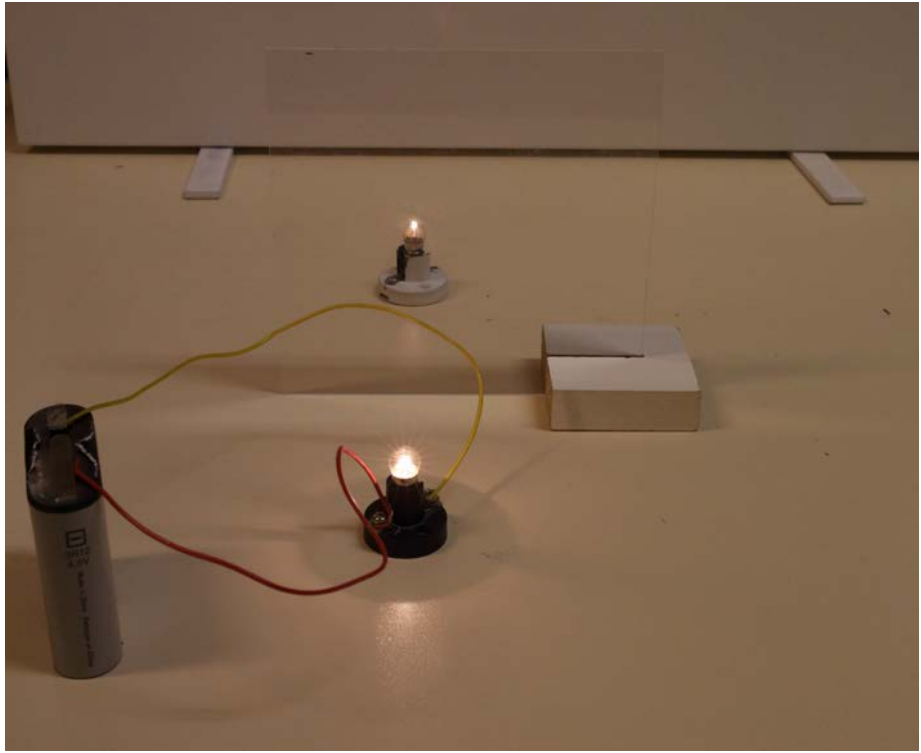

**a)**

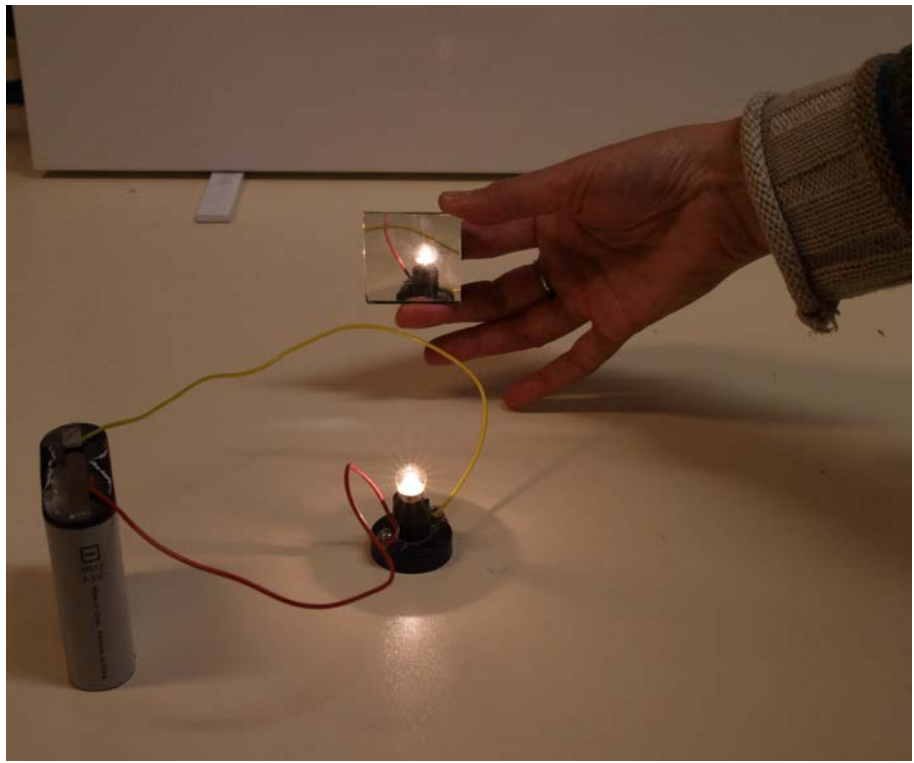

**b)**

**Fig 2: virtual images obtained by reflection : a) with a Plexiglas plate ; b) with a plane mirror**

*d) Focal distance of a thin lens*

<u>*Step 1:*</u> Pursuing with qualitative experiments, the students use a thin convergent lens (some of them have never used one before) and are asked to define the conditions for projecting an image of a vertical slit onto a screen. Lenses of different focal distances are given to the students. During this experiment that turns out to be more complex than it appears at first sight, students experiment two important aspects of image formation with a lens:

- the nature of the image obtained depends on the distance of the object to the lens: objects 'far' from the lens yield an image that can be projected onto a screen (a real image), whereas objects 'close' to the lens yield an image that can be seen by eye directly through the lens, but cannot be projected onto a screen (a virtual image)
- in the case when a real image of the object (slit) is formed onto a screen, its size, brightness and position varies with the lens used.

<u>*Step 2:*</u> The professor presents the notion of a thin lens, its characteristics and its conditions of use as a stigmatic system (the notions of stigmatism and Gauss conditions have been introduced earlier). The notions of focus and focal distances are introduced both for convergent and divergent lenses and the basic rules of ray tracing are explained. Emphasis is given on the fact that focus points are abstract, not material points of the lens, and on the fact that the image focus is *not* the image of the object focus, a common mistake for students meeting these notions for the first time.

<u>*Step 3:*</u> The students then lead three quantitative experiments aimed at measuring the focal distance of a lens. A white light source with a condenser lens is used to form a parallel beam (the parallelism is checked by moving a screen in the beam direction and by verifying that the beam section is constant). A convergent thin lens is placed in the beam and the convergent point is seeked for in the refracted beam. The lens – convergence point distance is measured and various sources of uncertainties are discussed. The second experiment proposed is the classical auto-collimation method: the image focal plane is folded onto the object focal plane of a lens using a small plane mirror after the lens. The image of an object is then formed onto the object plane itself. Students make a sketch of the set-up and trace several light rays originating from the object to conclude that the image forms indeed in the plane of the object. They obtain a new estimate of the focal distance by measuring the object-lens distance in this configuration. Finally, the students measure the positions of the image for various positions of the object and use the conjugation relationship of the lens to plot the inverse of the position of the image (with respect to the lens center) as a function of the inverse of the position of the object. The inverse of the intercept of this curve, taking into account the dispersion of the measurements, leads to an estimate of the focal distance and its uncertainty.

All three measurements are then compared with their uncertainties and the advantages and drawbacks of each method are discussed. The students are then asked which method is their favorite and why.

### 3. Originality, advantages and drawbacks

A key advantage of our approach is to make the students both more active and motivated in their studies. By making them experiment and answer questions by themselves, students become actors of their learning. The interaction among students is not only allowed but even strongly encouraged. The running of the practical sessions also implies a change in the role of the teacher and its relationship with the students during learning. The proximity between students and their teacher is developed, the teacher guiding the students in their learning rather than delivering a pre-conceived knowledge. Practical sessions are especially suited for discussions and personal follow-up.

One potential drawback of promoting this approach in a first year university course is that it requires a certain level of maturity from the students which some of them do not have at this level of their curriculum. For those students, this drawback is in practice partially compensated by the interaction they have with more mature students which helps them to adapt to this way of learning. The running of the sessions also requires a degree of maturity from the professors in order to break the habits acquired in more traditional courses. This could be a drawback but our experience shows that even young professors joining the pedagogical team adapt very well to the course philosophy and practice.

The specificity of our approach resides in the *combination* of various novel teaching techniques that may have been individually used in previous works but, to the best of our knowledge, not combined at the same time in a single course. Indeed, our course combines the following various aspects :

- putting experiment at the heart of the teaching process,
- giving the students an active role in learning through experiment *before* acquiring any theoretical knowledge, in the line of other 'active learning' experiments,
- shifting the role of the teacher from being an instructor to acting as an assistant to guide the students in their learning,
- mixing in a single session different activities that usually belong to separate sessions. Indeed, in our course, a single session mixes time for experiments, time for lectures and, in some cases, time for exercise solving. Typically, students and teacher switch among these activities several times in a given session.

Moreover, we have developed these techniques in a course devoted to geometrical optics, a field of physics traditionally taught in a quite formal way. Based on practicing this course for a decade, we argue that our approach has been beneficial both for teaching geometrical optics and introducing experimental physics to the students.

### 4. Associated teaching materials

Several pedagogical supports have been developed to complement the practical sessions described in section 2.

a) Handout

A detailed handout is provided to the students to guide them through the sessions, in the spirit of, e.g., the training manuals by [9]. The handout has been conceived as an experimental notebook in which the students are invited to record their observations and results. For each experiment, empty spaces are provided which the students fill in with their observations and schematics. For the quantitative experiments, tables with empty cells are provided for recording measurements. Measurements of a given quantity with different methods are synthesized in special summary boxes which provide a global view of the different results along with their uncertainties and give a feeling of which method is best suited, given the goal pursued (highest precision, easiest set-up, quick estimate, …). In some cases, the elements of an experiment are sketched for the students to complete as their experiments proceed (ray tracing, image positioning, …). In between the boxes, elements of the course are given so that the handout, mixing concepts and experimental results, can be read in a linear and coherent way. This is especially useful for the student when preparing their final examination (see section 6).

Another feature of the handout is to propose further reading sessions dealing with history and modern applications of geometrical optics. These topics are presented in specific inserts and are not necessarily developed during the sessions. The interested student thus finds material to develop his/her knowledge on topics such as:

- The corpuscular theory of light : Planck's and Einstein's quanta,
- Modern applications of optical fibers,
- A microscopic model for explaining the reflection of light,
- The Hubble Space Telescope,
- Some "non optical" images,
- Animal vision,
- Optical instruments,
- Different origins of colors,
- Lasers and their applications, …

In total, 23 such boxes are proposed in the handout. As an illustration, a typical page of the handout is presented in Fig. 3.

# 4. Light propagation in an inhomogeneous material

## 4.1 Experiment

<u>Experiment 11</u> : *Bending of a light ray*

*A 3mm layer of sugar is put into a tank. The tank is then slowly filled with water. After about an hour, a red laser beam is sent horizontally across the water.*

*Describe your observations. Make a drawing.*

## 4.2 Interpretation

The above experiment demonstrates that the principle according to which light propagates along straight lines is not always valid. It is true only in an *homogeneous* material. In an *inhomogeneous* material, light rays are bent. Let's try to understand why.

In an *inhomogeneous* material, the refractive index is not uniform : its value changes from one point of the material to another. Let's assume that this variation is slow enough: for example, in our experiment, the index slowly increases from 1,33 at the water surface to 1,5 (sugar saturated water) at the bottom of the tank.

With this assumption in mind, we can build a model in which the sugared water in the tank is split into thin horizontal layers. Each layer is thin enough for the index to be considered uniform in the layer to a good approximation. Between two contiguous layers, the index varies by a small amount $\delta n$ (positive when considering layers from top to bottom). The material can thus be modeled as a series of horizontal layers with different refractive indexes. A light ray incident on such a material will experience a series of refractions following Snell's law at the interfaces of successive layers (see figure below). As the index increases from top to bottom of the water, at each layer interface the light ray gets closer to the normal to the interface : it bends.

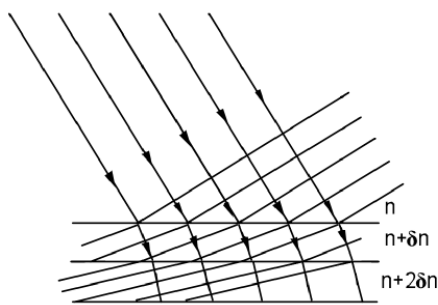

- This phenomenon explains the *mirages* one often sees during the summer on a heated road : the index variation here is due to the temperature varying with altitude.
- Light rays propagating in optical fibers with graded index are also bent.

**Fig 3 : example page of the handout**

b) Recipes for handling uncertainties

A rigorous treatment of uncertainties is clearly beyond the scope of a first year course (not mentioning the fact that most students do not have the mathematical background for that) but we feel that it is essential for students of an experimental science to grasp a sense of the uncertain nature of any measurement. This is why a two hour session at the beginning of the course is devoted to uncertainties in experimental physics. Uncertainties are introduced through the dispersion of the repeated measurements of a quantity, leaving aside systematic uncertainties. Namely, each student is asked to measure a given distance with a ruler. Results are collated by the professor and an interactive discussion is conducted on their dispersion, course, significant figures, and finally on expressing the result as a range in which the true value has 'great probability' to be found. This allows us to describe a measurement as a given realization of a random variable following a probability distribution. The problem of propagating uncertainties is also introduced, for quantities depending on one variable. A 'mathematical' recipe is given: the uncertainty on the dependent quantity is related to the uncertainty on the independent variable through the (absolute) derivative of the function relating the two quantities: the evaluated and the measured value. This recipe is often used in the experiments of the practical sessions. Examples in various fields of physics are taken to illustrate this concept (Ohm's law, Hooke's law, Snell's law). We do not attempt to generalize to the case of quantities depending on more than one measurable quantity as most students at this stage do not know partial derivatives and as most experiments of our course do not require it.

c) Electronic support

We have developed an electronic course accompanying the sessions available on the university electronic platform. It contains applets allowing students to 'practice', e.g., ray tracing, as well as course complements, quizzes for self-evaluation of their understanding and knowledge. A marker A (for Application) is indicated in the handout each time some electronic material is provided on the platform in relation to a notion or an experiment

d) Tutorial sessions

Parallel to the experimental sessions described in section 2, tutorial sessions take place (see Table 1 in the appendices). These sessions are conducted in a more traditional way: application exercises (also provided in the handout) are made by the students, then corrected by the teacher. Here are some examples of the topics treated in the exercises:

- Rain detector
- Cladded optical fiber
- Where does the fisherman see the fish?
- An model of the eye
- Correction of a nearsighted eye
- Rainbow
- Expansion speed of a galaxy measured by spectroscopy, etc.

The use of clickers has been introduced during these sessions to make the students active. Each professor defines his/her own clicker questions that can be asked at the beginning of the exercise to check the students intuition about the expected result, or at the end in order to check the number of students who solved the problem correctly, etc.

### 5. Organization of the course

This teaching is proposed to first-year students of two pathways: "Mathematics and Computer Science" and "Physics, Chemistry and Earth sciences". The total number of students per year is approximately 450. Both the number of students and the specificity of the course imply stringent organization. The principle is that for a given group of 20 students, all sessions are given by a single teacher. This ensures a continuity in the teaching and makes evaluation easier. It also softens the difficult transition from high school to university.

All practical sessions are given in devoted laboratories containing the required material for the experiments. Each laboratory is fully equipped. There are 3 different laboratories duplicated twice for a total of 6 rooms. The tutorial sessions are given in ordinary classrooms.

### 6. Assessments of the students work

The students are evaluated both by continuous assessment and by final exams. The continuous assessment is divided in individual lab reports (usually they are written by pairs of students) and short quizzes. The aim of continuous assessment is to force the students to work regularly and to ease the transition from high school, where they were not used to final exams as a unique assessment.

Two final exams are organized: a written one with problems to solve, plus a practical exam, which is quite unusual at university in France. During this practical exam, students do experiments and measurements similar to what they practiced during the practical sessions, and record their findings in a notebook. The assessment is based both on the practical skills of the student who is asked to present and justify the methods employed, and on the recorded results. The scores of students at this practical exam are usually good.

### 7. Feedback

The course in its present form has been proposed to first year students at Orsay for a decade, and its efficiency in terms of the goals pursued can be put into perspective thanks to the numerous feedbacks of both students and professors over the years.

Student feedback is made through a final anonymous survey, as for all the courses of the first-year. Feedbacks are very positive. The proposed learning is clearly helping them to acquire autonomy and, to some extent, allows them to work and assimilate notions at their own pace. Specifically, 90% of the students think that the organization of the teaching between practical sessions including course and tutorials is well adapted, and 70% find this teaching interesting. Regarding practical sessions, 80% find that the equilibrium between course and experiments is good. The complementarity of theoretical knowledge with practical sessions as well as with tutorials is good for 76% of the students. The handout is clear for 86% of them, and the further reading sessions are interesting for 80% of them.

The feedback from professors is also very positive; there are mainly young professors in this course (PhD students and young assistant professors) who are happy to have the opportunity to teach an entire and consistent course, including lectures (it's very rare for young professors to have the opportunity to lecture), tutorials and practical sessions.

## 8. Comparison with the classic approach

The ideas that make the substance of the new course described in our paper originated from teaching geometrical optics in a more classical way (in which lectures, practical and tutorial sessions are separated and in which practical sessions are often a mere verification of the course) and are based on our direct experience with students. Even if it is almost impossible to make in retrospect a quantitative comparison between the classical way of teaching and the one proposed in this paper, several key aspects can be qualitatively compared that support our new approach.

First, the motivation of the students has been dramatically improved with our new approach: teaching shifted from a mere and often tedious verification of a physics law to proceeding through the steps of a real and complete physics experiment. In our new approach, no *a priori* knowledge is required. Students just have to experience with the material given and are encouraged to discuss their observations and results with their neighbours. Motivation arises from various sides:
- being able to achieve the assigned experimental goal;
- being able to produce a visually satisfying result (e.g., a 'beautiful', undistorted image);
- being able to play at will with the experimental setup to modify or improve the result
- being the first ones to achieve the goal.

The recreational dimension of this approach (students are encouraged to 'play' with the material at will), that is absent from the traditional 'verify the law' approach, also contributes to the student motivation.

Second, with our approach, students improve their experimental skills. This is not only due to the fact that they practice more, but also and mainly to the fact that they are asked to be more innovative in their practice, compared to the classic approach. To put it in a nutshell, students are not following a recipe: they are inventing a recipe. This is one of the key concepts of our teaching. While the classic approach favours a more abstract approach of learning, our new approach focuses on a more direct and concrete learning aiming at developing practicing skills.

Third, memorization on the long term of the concepts taught has been qualitatively assessed by some of the teachers of this course. In parallel of the 1st year course discussed here, some teachers also taught 3[rd] year practical sessions in optics. Several students who had attended our 1[st] year course two or three years earlier followed these sessions. The qualitative feedback from watching these students practicing optics two years after our course is the following:

- basic experiments of general use for any optical layout were memorized (e.g., self-collimation, use of a lens, focal length determination). Students were able to rapidly determine the focal length of a given lens, and knew that self-collimation was a method to obtain a parallel beam (although the details of how to proceed were often forgotten and had to be explained again). This contrasts with students who followed a more classical way, in particular those coming from preparatory courses for Grandes Ecoles.
- The ability of manipulating simple optical devices was manifest. The students were not 'afraid' of experimenting. Obviously, they had gained in maturity in two years and some reflex in practicing optics seemed to be acquired.
- When asked, many students expressed their interest in our course and approach. As one of our students put it several years after the course: 'It was the only course in which experiment was truly entangled with theory, that was original. I never had in my curriculum such an opportunity to alternate as much practice and theory'.

### 9. Future improvements

Many improvements can be considered to further develop our approach and render it even more efficient. For the practical sessions, computers could be used to help students analyze their measurements: a ray tracing software could help students understanding light propagation through optical devices. We also foresee the use of 2D CCD cameras to introduce concepts of numerical photography such as numerical aperture, resolution, … For the tutorials, a problem-solving approach [10] could be developed to supplement the more traditional exercises. As an example of applying this to geometrical optics, students could be given a photograph of a known monument (e.g. the Eiffel tower) and infer from its characteristics and their own knowledge of the monument the main properties of the objective lens that took it.

### 10. Conclusion

We have developed a teaching course in geometrical optics for first-year students based on experimentation: the students learn through experimenting. Geometrical optics is taught in a modern and interactive way: students are encouraged to be more active than in a classical course, which helps them to learn more efficiently and eases the transition into university. After a decade of experience, our approach can be put into perspective with respect to more traditional ways of learning. Through the feedback we have obtained over the years, both from students and professors, it appears that this course has contributed to changing the opinion of students on a field of Physics that too often suffers from the prejudice of being old-fashioned and disconnected from every day life. The methods presented in this paper could with little effort be applied to other courses of experimental science teaching.

## Acknowledgments

The creation of this course benefited from the inputs of many people, be it professors or students. We would like to thank in particular Eric Arié, Claude Cabot and Laurence Maurines who devoted a substantial amount of their time to prepare the pedagogy introduced in this course. We are thankful to Gaël Latour and Marie Godard for their strong involvement in the recent development of the course. More generally, we would like to thank all the teachers who have been involved in this course for a decade for their enthusiasm and for the new ideas they braught and that today continue to develop this teaching. Finally, we would like to thank Michael Joyce and Anniina Salonen for a careful reading and useful comments on the manuscript.

## Dedication

This paper is dedicated to the memory of Daniel Beaufils, an associate professor at the university Paris Sud, who passed away in 2013. Daniel was strongly involved in building the course presented in this paper. He was at the origin of the electronic course associated with it, and his inputs were key in implementing the pedagogy of the course in the handout.

**Appendices**

1. Content of the course

The outline of the course is typical for a course on geometrical optics and is summarized by the title of this teaching: "Light, images and colors". The first chapter deals with reflection and refraction of light, the second one is about optical images, the third one about lenses and human vision; the fourth chapter concerns colors and spectroscopy. Finally, there is a last chapter in the handout devoted to measurement errors. The order of the handout is not exactly the one followed in the sessions since measurement errors are treated at the beginning (see Table 1 for the organization of the sessions), just after the first measurements. This way, the students understand the interest of error bars, and immediately apply them to these first measurements and to all the other ones in the course (and, we hope, in all other courses).

2. Organization of the sessions

| Session | Topic |
|---------|-------|
| S0 | **Lecture** : the nature of light<br>**Exercises** on the propagation of light |
| S1 | **Practical course** : reflection and refraction of light |
| S2 | **Lecture** : measurement errors<br>**Application** : measurements made at S1<br>**Exercises** : reflection and refraction |
| S3 | **Practical course** : mirrors and interfaces |
| S4 | **Exercises** : mirrors and interfaces |
| S5 | **Practical course** : lenses |
| S6 | **Exercises** : lenses |
| S7 | **Practical course** : human eye ; association of lenses |
| S8 | **Exercises** : lenses |
| S9 | **Practical course** : spectroscopy and colors |
| S10 | **Exercises** : spectroscopy and colors |
| S11 | **Practical exam** |

**Table1 : organization of the sessions**